\documentclass[nonacm,sigconf]{acmart}

\renewcommand\footnotetextcopyrightpermission[1]{} 

\setcopyright{none}

\AtBeginDocument{
    
}

\usepackage[normalem]{ulem}
\usepackage{xcolor}
\newcommand{\ph}[1]{\textcolor{black}{#1}}

\newcommand{\rt}[1]{}

\usepackage[most]{tcolorbox}

\newtcolorbox{takeaway}{
    colback=gray!5,           
    colframe=black!60!black,   
    arc=0mm,                  
    boxrule=0pt,              
    leftrule=4pt,             
    enhanced,                 
    breakable,                
    fontupper=\small\itshape, 
    before skip=10pt,         
    after skip=10pt           
}

\begin{document}

\title{MergirafSemi: A Language-Agnostic Semistructured Merge Tool}


\author{Pedro Lopes}
\affiliation{
  \institution{Centro de Informática, UFPE}
  \city{Recife}
  \country{Brazil}
}
\email{phls2@cin.ufpe.br}

\author{Paulo Borba}
\affiliation{
  \institution{Centro de Informática, UFPE}
  \city{Recife}
  \country{Brazil}
}
\email{phmb@cin.ufpe.br}

\author{Paola Accioly}
\affiliation{
  \institution{Centro de Informática, UFPE}
  \city{Recife}
  \country{Brazil}
}
\email{prga@cin.ufpe.br}

\author{Guilherme Cavalcanti}
\affiliation{
 \institution{Instituto Federal de Pernambuco}
 \city{Belo Jardim}
 \country{Brazil}
}
\email{guilherme.cavalcanti@belojardim.ifpe.edu.br}


\renewcommand{\shortauthors}{}
\renewcommand{\shorttitle}{}

\begin{abstract}
Developers frequently face merge conflicts when integrating concurrent changes. Most merge tools rely on unstructured, line-based comparisons, often producing spurious conflicts and missing actual conflicts. To address these limitations, structure-aware merge tools have been proposed, which leverage syntactic representations to improve merge accuracy. However, fully structured tools may incur higher computational cost, and language-specific tools require significant development and maintenance effort. To balance these trade-offs, we propose \textit{MergirafSemi}, a language-agnostic semistructured merge tool that captures structural information without requiring full structural modeling or language-specific implementations, \ph{and applies line-based merging within specific program regions, such as method bodies in Java}. Our tool leverages lightweight \ph{Concrete Syntax Trees} to guide merging decisions while preserving flexibility and efficiency across languages. We evaluate our tool through an empirical study on real-world merge scenarios across multiple programming languages, comparing it with unstructured, semistructured, and structured tools. Our results show that increasing structural granularity improves automatic conflict resolution but can also lead to more aggressive merge decisions, increasing the number of missed actual conflicts. In contrast, \textit{MergirafSemi} achieves a more balanced trade-off, reducing spurious conflicts while maintaining competitive accuracy and better runtime performance in most common scenarios. Compared to an unstructured tool, it substantially reduces spurious conflicts, and when compared to a semistructured language-specific tool, it achieves comparable effectiveness while exhibiting more robust execution behavior and significantly lower runtime overhead.
\end{abstract}


\frenchspacing
\maketitle

\section{Introduction}
\label{sec:introduction}

Developers frequently make changes to a codebase in parallel before merging their work. Version Control Systems (VCS) support this process by automatically merging different revisions of files. Most VCS mechanisms rely on unstructured, line-based comparison algorithms, such as \textit{diff3} \cite{diff3}, which operate solely at the textual level and do not account for the syntactic or semantic structure of the code \cite{formal_diff3}.

Although widely adopted and language-independent, line-based merging treats code as plain text, often producing spurious conflicts (false positives) that require unnecessary manual resolution of compatible changes. It may also fail to detect actual conflicts (false negatives), allowing syntactic or semantic inconsistencies to propagate into the merged artifact.

To address these limitations, structure-aware tools incorporate syntactic information into the merge process \cite{structured-tools-hunt,structured-tools-sven,structured-tools-spork,structured-tools-zhu-jdime,semistructured-tools-sepmerge,semistructured-tools-sven}. They represent code trees and perform \ph{node} matching and merging, improving accuracy over purely textual tools \cite{structure-is-good}. However, many of these tools rely on language-specific parsing, which increases development and maintenance effort, especially in multi-language projects.

More recently, language-agnostic structured tools, such as \textit{Mergiraf} \footnote{\textit{Mergiraf}'s Website: \url{https://mergiraf.org/}} and \textsc{LastMerge} \cite{structured-tools-lastmerge}, have been proposed to overcome this limitation by leveraging generic syntactic representations. While these tools improve generality, structured merging may incur higher computational cost and lead to overly aggressive merge decisions, resulting in more false negatives.

Rather than relying on full structural representations, semistructured tools use partial trees to guide merges and fall back to line-based merging when appropriate. This design reduces tree complexity while maintaining the benefits of structure, resulting in fewer false negatives \ph{and} more false positives than structured tools.

In this work, we build on this and introduce \textit{MergirafSemi}, a language-agnostic semistructured merge tool that captures structural information without requiring fully structured representations. It leverages lightweight Concrete Syntax Trees (CSTs) \cite{tree-sitter} and targeted line-based merging, enabling structure-aware decisions while \ph{maintaining accuracy}, flexibility, and simplifying support for new languages.

We evaluate our tool using a multilingual dataset of real-world, compilable merge scenarios across five languages (Java, JavaScript, Go, Python, and Rust). \ph{The} dataset construction and build validation methodology draws on prior work \cite{tools-evaluation-mike-ernst}, which proposed a similar pipeline exclusively for Java. Using this dataset, we compare our tool (\textit{MergirafSemi}) against unstructured (\textit{diff3}), structured (\textit{Mergiraf}), and language-specific semistructured (\textsc{S3M}) \cite{semistructured-tools-s3m} tools, as well as \textit{MergirafSemi+}, a variant with a different Java configuration for commutative contexts.

We assess the tools by measuring resolution rates, \ph{runtime performance}, and merge accuracy. To ensure accurate metrics, we compile and test tools' outputs, assessing false positives and false negatives through pairwise comparisons, specifically in scenarios where tools disagree on the existence of conflicts.

Our evaluation is structured around three research questions: (i) the trade-offs between structural granularity, accuracy, and runtime, (ii) how our tool compares to semistructured language-specific and unstructured tools, and (iii) the impact of changing commutative contexts on merge accuracy.

Our results yield three main insights. First, greater structural granularity improves conflict resolution but raises false negatives; \textit{MergirafSemi} offers a better balance among false positives, false negatives, and runtime. Second, it greatly reduces false positives compared to \textit{diff3} and matches \textsc{S3M}'s accuracy while running reliably (no timeouts or failures) with lower runtime overhead. Finally, relaxing commutative constraints in both tool versions has only a minor effect on conflict resolution, though it introduces a few false negatives \ph{in \textit{MergirafSemi+}}.

Overall, our findings suggest that semistructured merging can be effectively realized in a language-agnostic setting. Compared to unstructured tools, it substantially reduces false positives while reducing \ph{the} structured tools' high number of false negatives. This positions language-agnostic semistructured merging as a viable alternative to other language-agnostic tools, achieving competitive accuracy with language-specific semistructured tools.
\section{Unstructured, Semistructured and Structured Merge}
\label{sec:background}

To describe the limitations of unstructured merge, we present two representative scenarios in which line-based tools yield incorrect merge outcomes.

\begin{figure}[ht]
  \centering
  \includegraphics[width=0.75\linewidth]{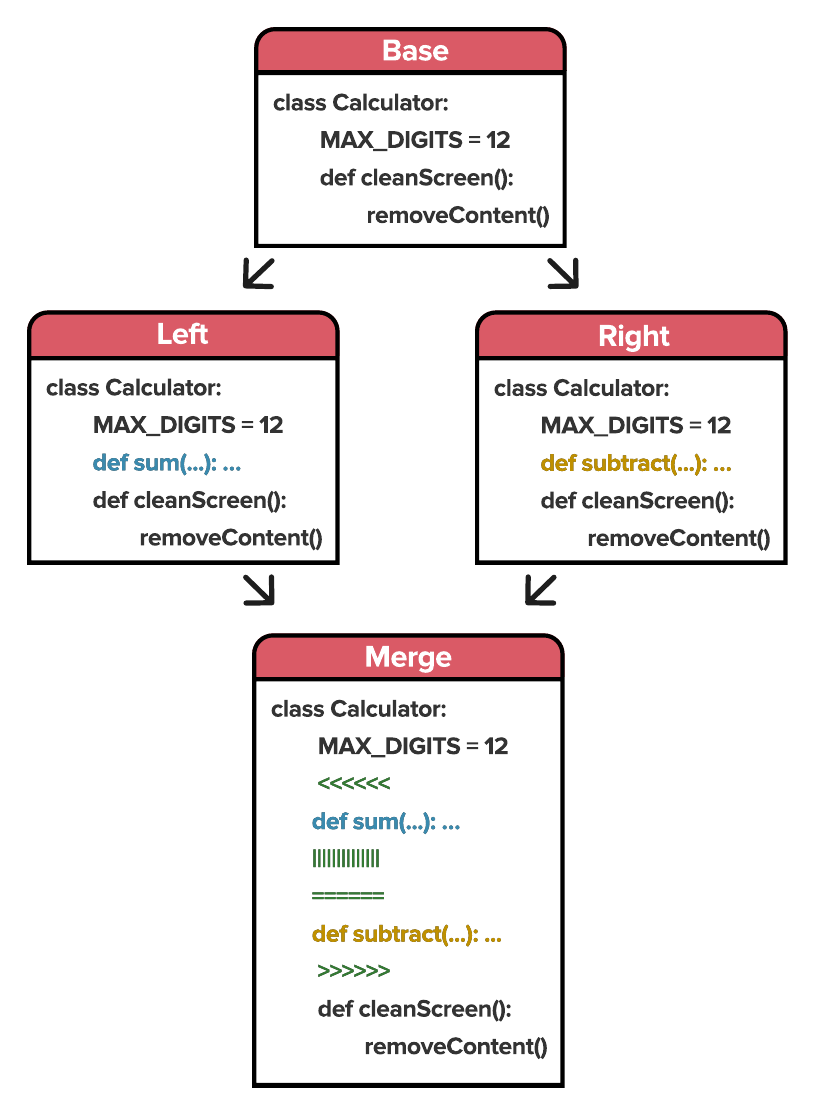}
  \caption{\label{fig:afpuns}Unstructured Merge False Positive.}
  \Description{An example of Unstructured Merge False Positive}
\end{figure}

Figure \ref{fig:afpuns} illustrates a scenario resulting in a spurious conflict. Two developers independently extend the same class by introducing new, semantically independent methods in the same textual region of the file. Because \ph{unstructured tools} compare overlapping line ranges \ph{against} a common base version, they interpret these edits as conflicting changes.

The merge output contains conflict markers that require manual resolution, even though the changes are compatible \ph{\cite{murta_conflict}}. This example constitutes a false positive: unstructured tools incorrectly signal a conflict \ph{even though there is no} semantic interference between the \ph{modifications}.

\begin{figure}[ht]
  \centering
  \includegraphics[width=0.75\linewidth]{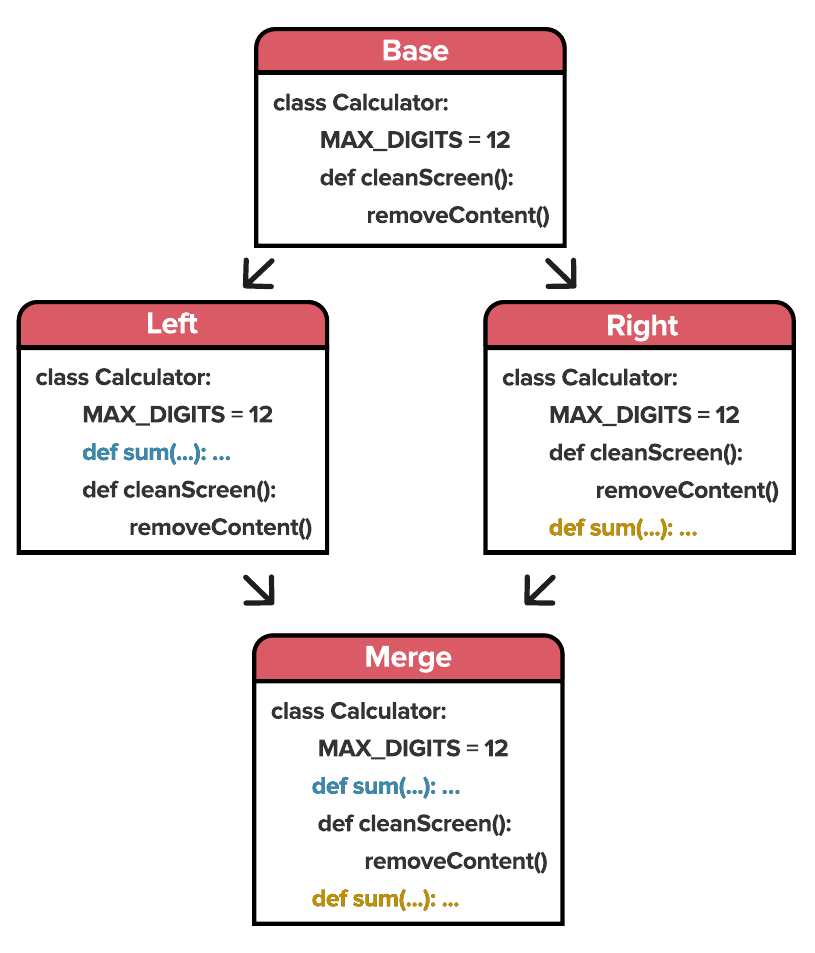}
  \caption{\label{fig:afnuns}Unstructured Merge False Negative.}
  \Description{An example of Unstructured Merge False Negative}
\end{figure}

In contrast, Figure \ref{fig:afnuns} presents a scenario that leads to an actual missed conflict. Two developers independently introduce a method with the same signature in different regions of the file. Again, as line-based merge detects conflicts based on overlapping lines \cite{formal_diff3}, it treats these edits as independent and combines them automatically.

Although no conflict is reported, both changes affect the same logical entity, potentially resulting in invalid or unintended behavior. In languages such as Python, this can lead to unintended overriding depending on declaration order. This outcome represents a false negative.

These examples highlight a key limitation of unstructured tools: decisions are based solely on textual alignment rather than the syntactic organization of the code.

Structured merge tools address this limitation by reasoning about program elements such as classes and methods \cite{semistructured-tools-sven}. They represent code as trees, commonly using Abstract Syntax Trees (ASTs), aligning corresponding elements across revisions and reporting conflicts only when incompatible changes affect the same node \cite{structured-tools-lastmerge}.

Semistructured tools use a partial structure: high-level elements are structured, while method bodies are treated as raw text. This hybrid representation improves accuracy without requiring full syntactic processing \cite{semistructured-evidence}.

Revisiting the examples from a structure-aware perspective highlights the effects of these tools. In Figure \ref{fig:afpuns}, where two developers add different methods to the same class, both structured and semistructured tools recognize that the changes involve distinct program elements. Thus, these modifications can be integrated automatically without conflict. In contrast, Figure \ref{fig:afnuns} shows two developers introducing methods with the same signature in different file sections. Here, these tools detect that both changes refer to the same program element, correctly reporting a conflict, preventing inconsistent or even invalid code.

However, structured and semistructured tools typically depend on language-specific parsers, syntactic representations, and merging engines, increasing implementation and maintenance effort, especially in multi-language environments. While semistructured tools reduce structural detail, they still depend on those too.

Recent work shows that structured merge can be made language-agnostic \cite{structured-tools-lastmerge}, but this has not yet been fully explored for semistructured merging. This motivates the tool proposed in this work.
\section{\textit{MergirafSemi}}
\label{sec:mergirafsemi}

As discussed in Sections \ref{sec:introduction} and \ref{sec:background}, structure-aware tools improve merge accuracy by reasoning about program elements, but introduce important trade-offs. Structured tools provide fine-grained analysis, reducing false positives, but may incur higher computational cost and lead to more false negatives. Semistructured tools reduce structural granularity, reducing false negatives but reporting more false positives; however, existing semistructured solutions still rely on language-specific representations, limiting their applicability.

Motivated by the limitations of unstructured merge tools and the trade-offs of structure-aware tools, this work builds on \textit{Mergiraf}, a prior language-agnostic structured merge tool, and introduces \textit{MergirafSemi}, a language-agnostic semistructured merge tool that combines semistructural reasoning with selective use of textual merging, preserving its language-agnostic nature through generic syntax tree representations while reducing structural granularity.

To support semistructured merging, we rebuild \textit{Mergiraf}'s representation and use of structural information during merging. In particular, we introduce (i) selective structural parsing, analyzing specific regions structurally, (ii) line-based merging on specific nodes, and (iii) adaptations to existing language profiles, tree-building algorithms, matching procedures, and new unification mechanisms to operate correctly under partial structural information.

Importantly, we design these modifications to be modular and configurable. The semistructured behavior can be enabled via configuration flag (\verb|--semistructured=diff3|) when desired, allowing the tool to revert to \textit{Mergiraf}'s original fully structured behavior. This design preserves backward compatibility and enables controlled experimentation with different merging tools.

Our tool aims to balance merge accuracy and practical applicability by leveraging lightweight structural representations through \textit{TreeSitter} parsers \cite{tree-sitter}, configurable language profiles, and a multi-phase execution pipeline, which we describe in the following sections.

\subsection{Generic Trees}
\label{subsec:trees}

\textit{MergirafSemi} represents source code using generic tree structures derived from parsing input files with \textit{Tree-sitter} \cite{tree-sitter}, a fast and extensible parser framework currently in production at GitHub. These structures capture syntactic elements such as classes, methods, and declarations, serving as one of the bases for structure-aware merging.

In \textit{Mergiraf}, all syntactic elements identified by the parser are represented explicitly as nodes in the tree. This enables the merge process to operate at a fine-grained level, including internal statements within method bodies.

In contrast, \textit{MergirafSemi} employs a partial representation of the program tree in the form of Concrete Syntax Trees (CSTs). Unlike ASTs, CSTs preserve syntactic elements such as delimiters and formatting. While high-level constructs such as classes and method signatures are preserved as nodes, selected regions of the tree are treated as unstructured text. These regions are determined by a language-specific configuration that specifies the node types whose internal content should not be further decomposed.

Consequently, subtrees corresponding to these node types are truncated and represented as leaf nodes containing raw text, which we call unstructured merge later on in the merge process. This design enables the tool to retain structural awareness where it is most beneficial, while avoiding the complexity of fully structured analysis in regions where it is less critical.

\subsection{Multi-Phase Merge Pipeline}
\label{subsec:merge_pipeline}

\textit{MergirafSemi} operates through a three-phase pipeline that incrementally increases merge granularity.

In the first phase, the tool performs an autotuned unstructured merge using a \textit{diff3}-style algorithm. If the result is conflict-free, it is then parsed to detect duplicated signatures. If neither conflicts nor duplicated signatures are found, the merge is accepted, avoiding further analysis.

If conflicts or duplicated signatures are detected, the tool advances to a second phase, where semistructured merging is applied locally, parsing only conflicting regions into CSTs. Then, node correspondences are computed using \textit{GumTree} \cite{gumtree} and an unstructured merge is applied on the raw text leaf nodes mentioned in Subsection \ref{subsec:trees}. Because GumTree primarily aligns nodes against the base revision, elements newly added in both revisions remain unmatched. To prevent duplicated code, it executes a unification step.

This unification step collects all unmatched nodes from both parents and evaluates their logical identities using signatures. If an unmatched node in the left revision shares the exact same signature as one in the right revision, they are unified into a single logical entity.

Lastly, if conflicts persist or tree-related issues, such as parsing errors, are detected from Phase 2, a third phase applies semistructured merging globally by parsing the entire files into CSTs and repeating the matching, merging, and unification steps described in Phase 2.

This design enables efficient handling of simple cases while still supporting semistructural reasoning when needed. \textit{Mergiraf} also shares this pipeline, but on Phases 2 and 3, instead of semistructured tree-building, matching, and the unification step (which is exclusive to \textit{MergirafSemi}), the whole process is fully structured.

\subsection{Language Profiles}
\label{subsec:lang_profiles}

A key aspect of \textit{MergirafSemi} is the use of configurable language profiles, which define how structural information is interpreted and used during merging. These profiles enable the tool to remain language-agnostic while still leveraging language-specific insights.

Specifically, language profiles define signatures, which capture the essential attributes of program elements (e.g., function names and parameters) and are used to identify corresponding nodes across revisions. They also specify commutative contexts, where the order of elements does not affect code semantics, allowing the merge process to integrate them without conflicts.

Most importantly, the language profile controls the level of structural granularity through tree-truncation rules. These rules determine which node types should be treated as text leaves. When such nodes are encountered, their internal contents are treated as raw text and merged using unstructured tools.

This lightweight and configurable design enables \textit{MergirafSemi} to support multiple programming languages while significantly reducing the effort required for extension. In practice, adding support for a new language typically requires only a few hours, depending on the availability of an existing \textit{Tree-sitter} parser and the developer’s familiarity with the language. When structured support is already available within \textit{Mergiraf}, the configuration effort is almost minimal, since you only need to configure the nodes where the CST will be truncated. When no prior support exists, adding a new language requires integrating a \textit{Tree-sitter} parser and validating the merge behavior of the configurations cited in the previous paragraphs, but this effort is still relatively low compared to that required for language-specific tools.

\subsection{Structure-Aware Merging}
\label{subsec:merge_strategy}

During the semistructured phases of the pipeline, our tool performs merging based on the alignment of syntactic nodes across the Base, Left, and Right revisions (as illustrated in Section~\ref{sec:background}), as determined by \textit{GumTree} \cite{gumtree}.

When corresponding nodes are modified independently in compatible ways, such as changes to different attributes of the same construct or identical modifications, the tool automatically integrates the changes, as described in Section \ref{sec:background} and illustrated in Figure \ref{fig:afpuns}. Similarly, modifications affecting different child nodes of a common parent are merged without conflict.

Conflicts are reported when concurrent changes affect the same aligned node in incompatible ways and no deterministic resolution can be inferred, as described in Section \ref{sec:background} and illustrated in Figure \ref{fig:afnuns}. In such cases, the merge process preserves both versions and explicitly signals the conflict.

This applies only to the structured portion of the representation. When truncated nodes are encountered, their contents are merged using unstructured tools, enabling the tool to distinguish between independent and interfering changes based on program structure, rather than relying solely on textual overlap.

\subsection{Role of Textual Merging}
\label{subsec:textual_merging}

Textual merging complements structural analysis at various stages of the \textit{MergirafSemi} pipeline.

In the initial phase, the tool relies on line-based merging to efficiently handle simple scenarios. This fast-path mechanism, also adopted by language-specific structured merge tools such as \textit{jDime}~\cite{structured-tools-sven}, avoids the overhead of structural analysis when changes can be safely integrated using unstructured merging.

In semistructured phases 2 and 3, textual merging is applied within regions intentionally treated as unstructured, as defined by the language profile. For example, method bodies may be represented as plain text, enabling the tool to reuse line-based merging in contexts where fine-grained structural reasoning is unnecessary.

Finally, textual merging serves as a fallback during structure-aware phases. When the structural merge process cannot reliably resolve a specific node, the tool locally reverts to line-based merging for that region while preserving the surrounding structural context.
\section{Evaluation}
\label{sec:evaluation}

To evaluate \textit{MergirafSemi}, we performed an empirical study involving merge scenarios drawn from real-world projects across multiple programming languages. We aim to measure and compare its merge accuracy and runtime performance against unstructured (\textit{diff3}), fully structured (\textit{Mergiraf}), language-specific semistructured (\textsc{S3M} \cite{semistructured-tools-s3m}), and \textit{MergirafSemi+}, a variant of our tool with a different Java configuration within commutative contexts.

Our analysis centers on cases where different merge tools yield different results, allowing us to isolate instances in which tool selection directly influences the merge outcome.

\subsection{Research Questions}
\label{subsec:rqs}

We addressed the following Research Questions (RQs) to evaluate the effectiveness of \textit{MergirafSemi}. The third RQ additionally considers \textit{MergirafSemi+}.

\subsubsection{\textbf{RQ1}: What is the impact of structural granularity on merge accuracy?}
\label{subsubsec:rq1}
\hfill\

To investigate this question, we compare \textit{MergirafSemi} with the fully structured tool, \textit{Mergiraf}.

Fully structured merging relies on detailed syntax trees to align and integrate program elements across revisions. Semistructured merging reduces this granularity by treating selected regions of the code as text, allowing the merge process to rely on unstructured tools in contexts where fine-grained structure may not be necessary.

This comparison evaluates how structural granularity affects merge accuracy and whether reducing it helps avoid false positives or increases the likelihood of false negatives. 

\subsubsection{\textbf{RQ2}: How does a language-agnostic semistructured merge tool compare to language-specific and line-based tools?}
\label{subsubsec:rq2}
\hfill\

To address this question, we compare \textit{MergirafSemi} against two baselines: \textsc{S3M}, a language-specific semistructured merge tool for Java, and \textit{diff3}, a widely used line-based merge tool. The first comparison is restricted to Java scenarios, as \textsc{S3M} supports only Java, limiting cross-language evaluation.

Semistructured language-specific tools, such as \textsc{S3M}, leverage detailed knowledge of a target language to guide merge decisions, thereby improving precision but requiring substantial implementation and maintenance effort, limiting applicability across languages. In contrast, \textit{MergirafSemi} adopts a language-agnostic design, relying on generic parsing and configurable language profiles rather than language-specific logic.

This comparison evaluates whether a language-agnostic tool can achieve accuracy comparable to that of a language-specific one, without significant impact on merge runtime performance and accuracy, while also assessing its advantages over traditional line-based merging. In particular, we investigate whether removing language-specific aspects compromises merge accuracy relative to a specialized tool.

\subsubsection{\textbf{RQ3}: What is the impact of the commutative configuration on the accuracy of \textit{MergirafSemi} and \textit{MergirafSemi+}?}
\label{subsubsec:rq3}
\hfill\

We compare \textit{MergirafSemi} and \textit{MergirafSemi+} based on their handling of commutative constraints for class body declarations. This analysis focuses on Java, where certain declarations are treated as non-commutative, unlike other language profiles that do not enforce this constraint.

In the Java default language profile, elements such as methods and variable declarations are treated as non-commutative. As a result, if Left adds a method and Right adds an attribute (or vice versa) to the same class body, even when the changes are semantically independent and do not compromise program correctness, the merge results in a conflict.

\textit{MergirafSemi+} relaxes this constraint by treating such elements as commutative, allowing the merge process to integrate them without reporting conflicts. This modification is motivated by the observation that, in Java, the relative ordering of these declarations does not affect program semantics, although it may impact code organization.

This comparison evaluates how relaxing commutative constraints affects merge accuracy and if it introduces trade-offs, such as less-structured output due to interleaved declarations.

\subsection{Metrics}
\label{subsec:metrics}

We evaluate merge tools based on merge accuracy and runtime performance. To assess merge accuracy, we analyze how different tools behave in scenarios where their outputs differ on whether conflicts exist; in these cases, the tool choice affects the merge outcome.

Our analysis of merge accuracy is based on pairwise comparisons between tools. For each scenario, we compare the outputs produced by different tools and examine how they differ. These comparisons allow us to identify whether a tool introduced an unnecessary conflict or failed to detect an actual conflict.

We reference the merge results in the repository as the expected outcome, as they were produced and accepted by developers. However, since a merge can be correct even if it differs from the repository, we also conduct builds and tests for further verification.

We employ a standard methodology from recent empirical studies on merge tools \cite{tools-evaluation-mike-ernst, structured-tools-lastmerge}, combining repository comparisons with behavioral validation. By integrating various signals—such as syntactic equivalence, build success, and test outcomes—we obtain a more reliable approximation of the expected merge result, thereby minimizing misclassification risk.

\subsubsection{Automatic Conflict Resolution Rate}
\label{subsubsec:resolution_rate}
\hfill\

The automatic conflict resolution rate shows the percentage of merge scenarios a tool successfully resolves. We calculate this by dividing the number of conflict-free merges by the total valid scenarios. It serves as an initial indicator of the tool's ability to autonomously integrate changes, thereby reducing manual effort. However, a high rate mainly reflects conservativeness rather than accuracy, as it may merge code with errors. We interpret this alongside false positives and negatives (explained in the next section) to avoid hidden errors.

\subsubsection{Added False Positives and Added False Negatives}
\label{subsubsec:afp_afn}
\hfill\

We identify added false positives (aFPs) and added false negatives (aFNs) through pairwise comparison of tool outputs, following prior work \cite{structured-tools-lastmerge}. Given a pair of tools A and B, we say that A incurs an aFP candidate if it reports a conflict in a scenario where B produces a conflict-free result, and B incurs an aFN candidate if it produces a conflict-free result in a scenario where A reports a conflict.

To validate these cases, we rely on the repository merge as a reference and complement it with build and test execution. For aFPs, we require that B's output is either syntactically identical to the repository merge or, if different, it successfully compiles and passes all tests. For aFNs, we require that B's output differs from the repository merge and fails to compile or causes test failures.

These definitions capture two types of undesirable behavior: spurious conflicts (aFPs), which introduce avoidable manual effort, and silent integration of actual conflicts (aFNs), where incorrect merged artifacts are produced without signaling a conflict.

\subsubsection{Runtime Performance}
\label{subsubsec:runtime}
\hfill\

To assess runtime performance, we measure the execution time of all merge tools for each merge scenario.

Each scenario is executed six times per tool. The first execution is discarded as a warm-up run to mitigate initialization overhead (e.g., JVM startup in Java projects). The reported execution time corresponds to the mean of the remaining five runs.

To provide a comprehensive view of runtime behavior, we also analyze the distribution of execution times using rainclouds, which capture variability, median values, and the presence of outliers across scenarios.

\subsection{Experimental Setup}
\label{subsec:experiment}

This section describes the tools, execution environment, and experimental procedures used to evaluate the proposed tool.

\subsubsection{Merge Tools}
\label{subsubsec:tools}
\hfill\

We evaluate five merge tools: \textit{Mergiraf}, \textit{MergirafSemi}, \textit{MergirafSemi+}, \textsc{S3M}, and \textit{diff3}. The Mergiraf-based tools rely on version 0.12.1 of the original implementation, with \textit{MergirafSemi} and \textit{MergirafSemi+} built on top of it. For \textit{diff3}, we use the native implementation provided by the \texttt{git merge-file} command.

\textsc{S3M} does not provide strict versioning. Therefore, we use a specific version identified by a commit hash\footnote{\textsc{S3M}'s version via commit: \url{https://github.com/guilhermejccavalcanti/s3m/tree/bcf0a43cbc14d5aa21757686737c2434084e2165}}. To ensure reproducibility, we provide a replication package containing the exact binaries of all evaluated tools, along with the execution scripts and the complete dataset used in our experiments \cite{replication_package}.

\ph{Our baseline tool choices for each RQ are guided by the specific variable being controlled. For RQ1, we select \textit{Mergiraf} as the structured baseline because it shares an algorithmic foundation with \textit{MergirafSemi}. This allows us to attribute differences in accuracy and runtime to structural granularity rather than implementation choices. Using other structured tools like \textsc{LastMerge} \cite{structured-tools-lastmerge}, \textit{jDime} \cite{structured-tools-sven}, or \textit{Spork} \cite{structured-tools-spork} would introduce confounding factors related to heuristics and tree-building.} 

\ph{For RQ2, we choose \textsc{S3M} \cite{semistructured-tools-s3m} as the language-specific semistructured baseline, as it shares \textit{MergirafSemi}'s principle of combining structural and textual strategies, differing primarily in language-agnosticism. Tools like \textsc{SESAME} \cite{semistructured-tools-sesame}, which use syntactic separators, would not address RQ2 appropriately.}

\subsubsection{Execution Environment}
\label{subsubsec:environment}
\hfill\

All experiments are executed within Docker containers, using the \texttt{eclipse-temurin:17-jre-focal} image to standardize the runtime environment and Java dependencies (needed to execute S3M). The host machine runs Ubuntu 20.04.2 LTS (Kernel 5.4.0-74-generic x86\_64) and is equipped with an Intel Xeon E31220 @3.10GHz processor and 32GB of RAM.

To prevent anomalous executions from blocking the evaluation pipeline, we set a 30-minute timeout for both the merge process and the subsequent build and test step.

\subsubsection{Execution Pipeline}
\label{subsubsec:experiment_pipeline}
\hfill\

The evaluation uses a framework that clones repositories, filters merge scenarios, and isolates revisions. It is extended with custom components to run merge tools, normalize outputs, and perform build and test validation \cite{miningframework}. We normalize merge outputs to enable comparison by removing formatting differences, such as spaces, tabs, and line breaks introduced by pretty-printing in tree-based tools.

For each selected scenario, we extract the corresponding revisions into a clean environment and execute all tools via command-line interfaces.

Normalized outputs are compared to identify disagreements, and builds and tests are triggered following a validation procedure discussed in the previous sections. This is done using GitHub Actions with language-specific workflows.

To prevent biases introduced by project configurations such as linters and dependencies, we use generic workflows tailored to each language's build system. This ensures failures are due to issues in merged outputs, such as compilation or testing errors, rather than unrelated constraints.

\subsubsection{Dataset}
\label{subsubsec:dataset}
\hfill\

The dataset construction in this study is inspired by prior work on large-scale evaluation of merge tools \cite{tools-evaluation-mike-ernst}. While that study focused on Java repositories from Reaper \cite{reaper} and GitHub Greatest Hits \cite{github_greatest_hits}, our goal requires a multilingual setup including Java, JavaScript, Python, Go, and Rust.

Due to limited language support in Reaper, we rely solely on repositories from the GitHub Greatest Hits dataset, which are popular and actively maintained. We apply a multi-stage filtering pipeline to ensure the selected merge scenarios are suitable for evaluation.

Initially, we keep repositories containing build or dependency management files at the root, such as \texttt{pom.xml}, \texttt{build.gradle}, \texttt{package.json}, \texttt{Cargo.toml}, or \texttt{go.mod}, ensuring they are automatically compilable and testable.

We find the earliest successful build commit to set a stable baseline for merge evaluation. From there, we collect non-fast-forward merge commits where both parents modify the same files, ensuring meaningful merge decisions.

This filtering process yields a dataset of real-world merge scenarios that are both structurally relevant and empirically testable. In total, our dataset comprises \textbf{21,615} merge scenarios extracted from \textbf{513} repositories.

The distribution across languages is as follows: \textbf{7,181} scenarios from Go, \textbf{1,479} from Java, \textbf{2,842} from JavaScript, \textbf{5,860} from Python, and \textbf{4,253} from Rust.

By ensuring that all selected scenarios originate from buildable project states, we enable reliable validation of the expected merge result through compilation and test execution.
\section{Results and Discussion}
\label{sec:results_and_discussion}

In this section, we discuss the findings from our evaluation of \textit{MergirafSemi}. We focus on merge accuracy and runtime performance, highlighting how design choices affect outcomes. The results are organized by research questions, covering structural granularity, comparisons with existing merge tools, and commutative configurations.

\subsection{RQ1: What is the impact of structural granularity on merge accuracy and runtime?}
\label{subsec:result_rq2}

\begin{table*}
\caption{\textit{MergirafSemi} vs \textit{Mergiraf} -- Merge Accuracy}
\label{tab:rq1}
\begin{tabular}{lccccccc}
\toprule
 & & \multicolumn{3}{c}{\textit{\textbf{MergirafSemi}}} & \multicolumn{3}{c}{\textit{\textbf{Mergiraf}}} \\
\cmidrule(lr){3-5} \cmidrule(lr){6-8}
\textbf{Language} & \textbf{Total Scenarios} & \textbf{Res. Rate} & \textbf{aFPs} & \textbf{aFNs} & \textbf{Res. Rate} & \textbf{aFPs} & \textbf{aFNs} \\
\midrule
\textbf{Go}         & 7181 & 76,66\% & 153 & 65 & 79,81\% & 5  & 143 \\
\textbf{Java}       & 1479 & 81,27\% & 44  & 15 & 85,67\% & 0  & 36  \\
\textbf{JavaScript} & 2842 & 82,51\% & 61  & 5  & 86,14\% & 4  & 53  \\
\textbf{Python}     & 5860 & 79,88\% & 244 & 19 & 84,78\% & 10 & 90  \\
\textbf{Rust}       & 4253 & 83,92\% & 144 & 26 & 88,38\% & 9  & 81  \\
\bottomrule
\end{tabular}
\end{table*}

To answer this question, we compare \textit{MergirafSemi} (language-agnostic semistructured) with \textit{Mergiraf} (language-agnostic structured) and analyze their behavior across multiple programming languages.

\begin{figure}[h!]
    \centering
    \includegraphics[width=0.9\linewidth]{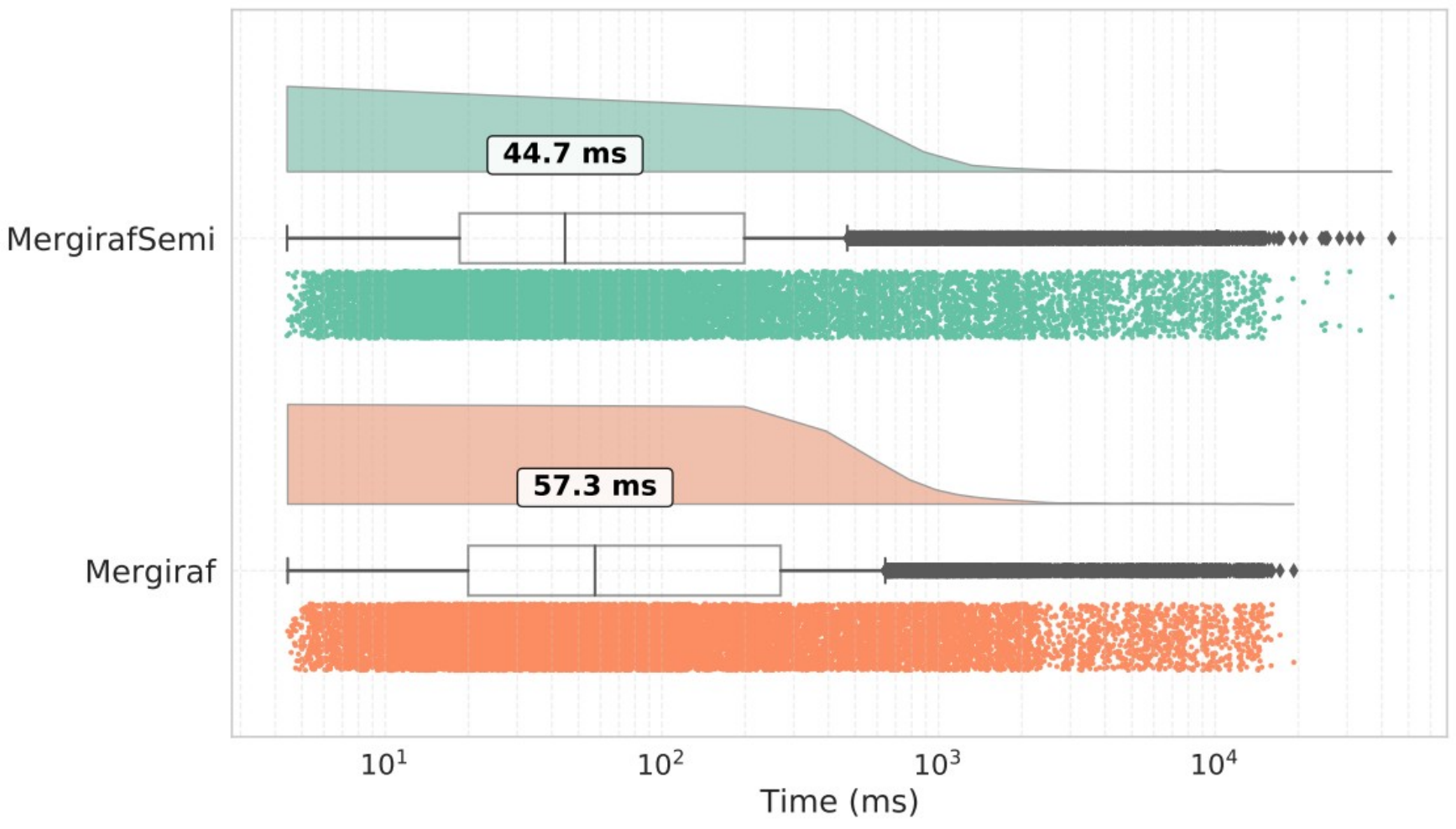}
    \caption{\label{fig:runtime_efficiency_rq2} Runtime Efficiency Across Programming Languages}
    \Description{A boxplot graph that shows runtime efficiency data of MergirafSemi and Mergiraf across multiple languages}
\end{figure}

Figure \ref{fig:runtime_efficiency_rq2} presents the aggregated runtime distribution on a logarithmic scale. The global median execution time shows that \textit{MergirafSemi} (44.7 ms) is 22\% faster than \textit{Mergiraf} (57.3 ms), indicating better performance in most common scenarios. Both tools also have many extreme outliers, with some cases exceeding $10^4$ ms.

A closer look at pipeline logs shows that during structural analysis phases, \textit{MergirafSemi} significantly reduces runtime. In Phases 2 and 3, the tree-building steps, semistructured merging is at least 60\% faster than structured merging. For example, in Phase 2 (local conflict resolution), median execution time in Rust drops from 781.9 ms to 89.5 ms, and in Java from 153.4 ms to 25.7 ms. Similar improvements occur in Phase 3, with \textit{MergirafSemi} reducing median runtime by about 61\% in Python and 63\% in Go.

This performance advantage comes at the cost of increased variability. While median runtimes are lower, average runtimes are affected by a few expensive scenarios, especially in Phase 3 of the merge process. This variability mainly stems from the unify operation (described in Section \ref{subsec:merge_pipeline}), which is costly in large files with many declarations due to the increased number of candidate comparisons. Sometimes, unify accounts for over 90\% of total time, impacting mean runtime and standard deviation.

As shown in Table \ref{tab:rq1},  across all evaluated languages, \textit{Mergiraf} resolves 84.1\% of merge scenarios without conflicts, outperforming \textit{MergirafSemi} by 4.1\% (80.0\%). Java’s success rate rises from 81.27\% to 85.67\%, Python’s from 79.88\% to 84.78\%. Similar gains are seen in Go, JavaScript, and Rust. These results indicate that structured merging is better at automatically resolving conflicts. \textit{Mergiraf} uses complete structural information to integrate concurrent changes that semistructured merging cannot, often completing in Phase 2 and avoiding further processing.

For merge accuracy, we examine aFPs and aFNs to clarify trade-offs. Fully structured merging sharply reduces false positives in all languages. For example, in Python, aFPs drop from 224 in \textit{MergirafSemi} to 10 in \textit{Mergiraf}, with similar trends in other languages. This shows fine-grained structure helps precisely identify independent changes.

However, fewer aFPs come with more aFNs: \textit{Mergiraf} often merges real conflicts without flagging them. For instance, in Go, aFNs rise from 65 to 143, and in Python, from 19 to 90.

This indicates that the structured tool’s higher resolution rate results from aggressive merging, which silently resolves conflicts. Structured merging reduces aFPs and boosts automation but risks more aFNs, while semistructured merging is conservative: it reports more aFPs but avoids many aFNs.

\begin{takeaway}
    \ph{\textbf{Answer to RQ1:} Neither tool is strictly better; each reflects different priorities. Structured merging favors automation and less manual effort, while semistructured merging prioritizes safer merges by preserving potential conflicts. Higher structural granularity improves conflict resolution and reduces aFPs, but doesn’t guarantee better correctness or runtime.}
\end{takeaway}

\subsection{RQ2: How does \textit{MergirafSemi} compare to a language-specific semistructured tool and a line-based merge tool?}
\label{subsec:result_rq2}

To answer this question, we compare \textit{MergirafSemi} (language-agnostic semistructured) against \textsc{S3M} (language-specific semistructured), focusing only on Java scenarios due to the language-specific nature of \textsc{S3M}, and against \textit{diff3} (unstructured) across all languages.

\begin{figure}[h!]
\centering
\includegraphics[width=0.85\linewidth]{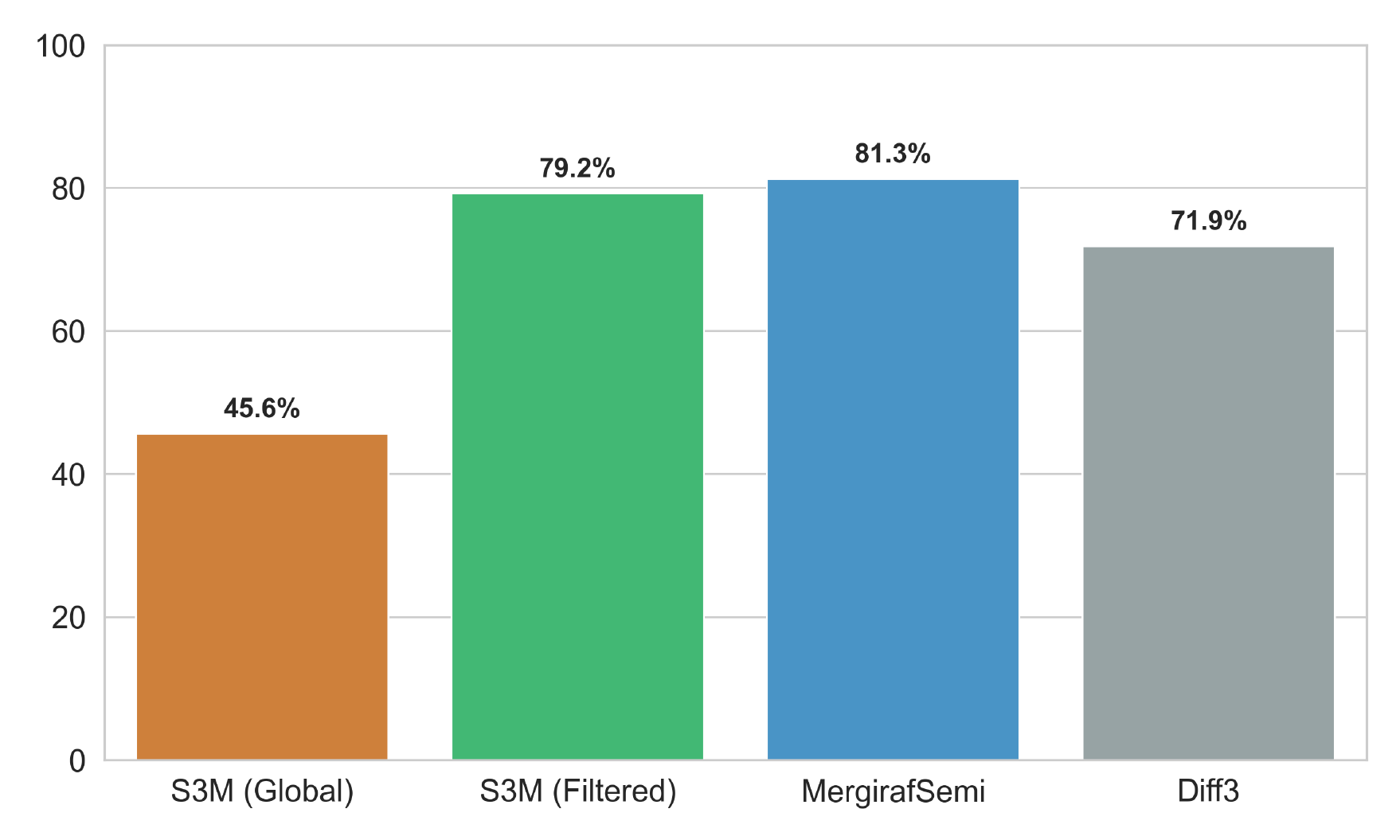}
\caption{\label{fig:resolution_rate_rq2} Automatic Conflict Resolution Rate (Java only).}
\Description{A bar graph comparing resolution rates of MergirafSemi, S3M, and diff3 in Java}
\end{figure}

In Figure \ref{fig:resolution_rate_rq2}, \textit{MergirafSemi} achieves a resolution rate of 81.3\%, outperforming both \textit{diff3} and \textsc{S3M} under default conditions. \textsc{S3M}'s initial global resolution rate of 45.6\% is strongly influenced by execution interruptions rather than merge effectiveness alone: it failed to generate a merge in 627 of 1,479 scenarios (approximately 42\%) due to timeouts and runtime errors.

\begin{figure}[ht]
\centering
\includegraphics[width=0.90\linewidth]{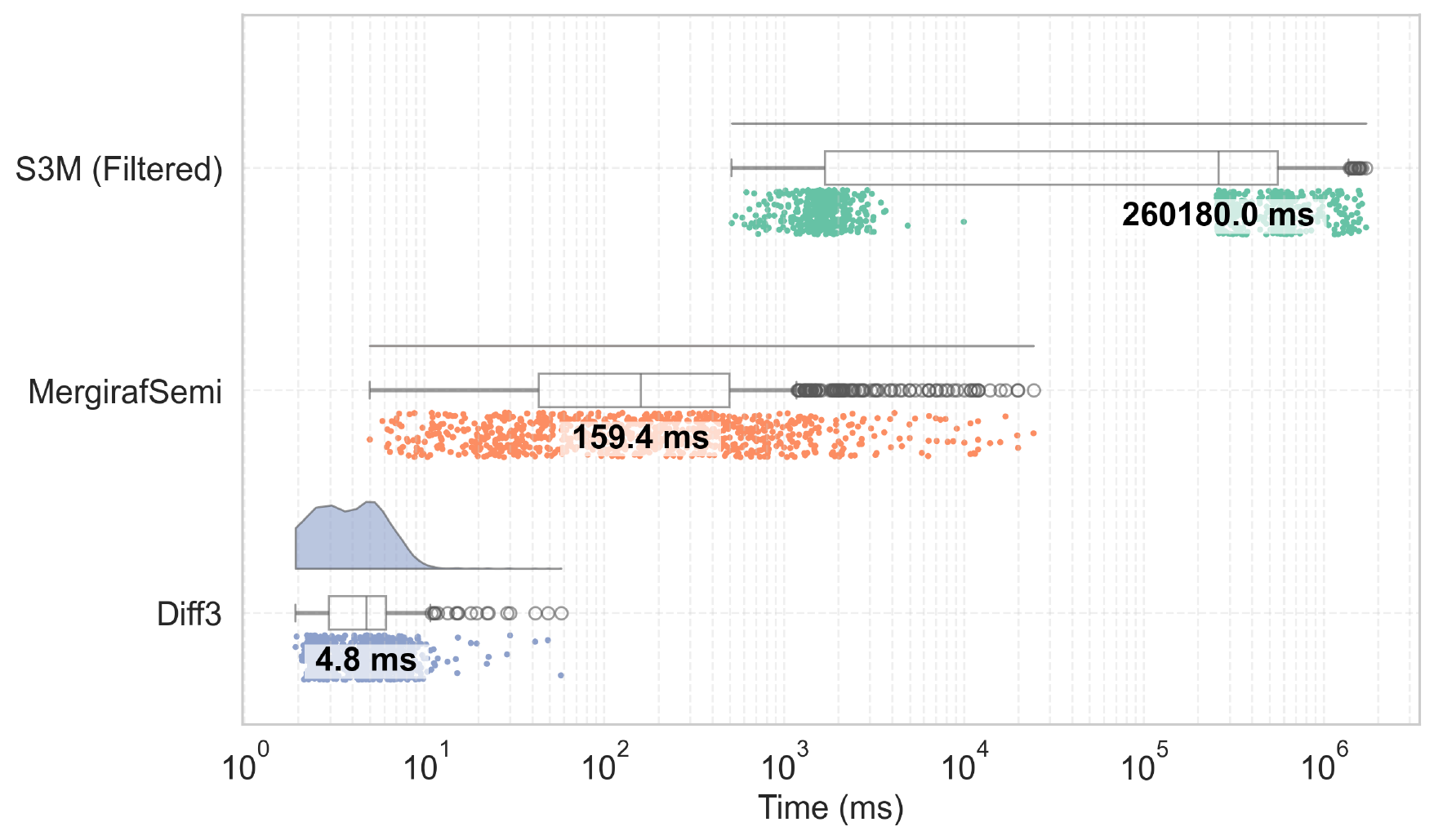}
\caption{\label{fig:runtime_rq2} Runtime Comparison Across Merge Tools (Java only).}
\Description{A boxplot comparing runtime distributions of MergirafSemi, S3M, and diff3}
\end{figure}

To better evaluate the tool, we now restrict the analysis to the 852 scenarios in which \textsc{S3M} completes successfully. Under these conditions, its resolution rate increases to 79.2\%, approaching the performance of \textit{MergirafSemi}, though still slightly lower. 

Despite restricting our analysis to the mentioned scenarios, \textit{MergirafSemi} successfully processed all 1479 scenarios without execution failures under the same experimental setup.

Figure \ref{fig:runtime_rq2} shows the runtime distribution of the evaluated tools. As expected, \textit{diff3} has the lowest median execution time at approximately 4.8 ms, reflecting its textual nature. In contrast, \textit{MergirafSemi} has a median runtime of 159.4 ms due to the overhead from parsing and structure-aware processing. Even in optimal scenarios, our tool incurs runtime overhead compared to standalone \textit{diff3} due to the parsing step required to detect duplicated signatures.

Even when considering only scenarios where \textsc{S3M} completes successfully, its median execution time reaches 260,180.0 ms (approximately 4.3 minutes) per merge. However, this comparison should be interpreted with caution, as the tools are implemented in different programming languages and runtime environments, which can significantly affect execution time.

Rather than focusing on absolute runtime differences, our goal is to assess whether a language-agnostic semistructured tool imposes a prohibitive runtime overhead. In this respect, the results indicate that \textit{MergirafSemi} remains efficient in practice, with execution times that are compatible with real-world usage.

\begin{table}[h]
\caption{\textit{MergirafSemi} vs \textsc{S3M} -- Merge performance}
\label{tab:rq2}
\begin{tabular}{ccc}
\toprule
\textbf{Language}              & \textbf{aFPs} & \textbf{aFNs} \\
\midrule
\textit{\textbf{MergirafSemi}} & 6             & 27            \\
\textbf{S3M}                   & 11            & 13            \\
\bottomrule
\end{tabular}
\end{table}

A deeper analysis of aFPs and aFNs offers insight into these tools' behavior. Table 
\ref{tab:rq2} compares the two semistructured tools across scenarios in which \textsc{S3M} completes successfully, showing a trade-off between conflict reduction and merge conservativeness.

\textit{MergirafSemi} has fewer false positives (6 vs. 11), indicating an ability to avoid unnecessary conflicts, while \textsc{S3M} has fewer false negatives (13 vs. 27), indicating its language-specific design favors conservative conflict handling. Although differences are significant, they should be interpreted cautiously, as implementation, configuration, and heuristics may influence outcomes, requiring manual analysis to clarify the specific reasons for these discrepancies.

\begin{table}[h]
\caption{\textit{MergirafSemi} vs \textit{diff3} - Merge Accuracy}
\label{tab:rq2_afps_afns}
\begin{tabular}{lcccc}
\toprule
                    & \multicolumn{2}{c}{\textit{\textbf{MergirafSemi}}} & \multicolumn{2}{c}{\textit{\textbf{diff3}}} \\
\cmidrule(lr){2-3} \cmidrule(lr){4-5}
\textbf{Language}   & \textbf{aFPs} & \textbf{aFNs} & \textbf{aFPs} & \textbf{aFNs} \\
\midrule
\textbf{Go}         & 20 & 292 & 65 & 9  \\
\textbf{Java}       & 3  & 110 & 33 & 1  \\
\textbf{JavaScript} & 45 & 44  & 89 & 1  \\
\textbf{Python}     & 1  & 172 & 30 & 44 \\
\textbf{Rust}       & 0  & 244 & 17 & 15 \\
\bottomrule
\end{tabular}
\end{table}

When comparing \textit{MergirafSemi} with \textit{diff3} across all languages, our tool significantly reduces aFPs by utilizing structural information, but it also leads to a notable increase in aFNs. While \textit{diff3} has few aFNs in most languages (except Python), \textit{MergirafSemi} generates many more, indicating that automating conflict resolution raises the risk of incorrectly merging changes without notifying developers.

\ph{Please note that \textit{MergirafSemi} has 6 aFPs when compared to \textsc{S3M} and 3 when compared to \textit{diff3}. This difference is expected because these metrics are computed through pairwise comparisons, as explained in Subsection \ref{subsubsec:afp_afn}. The set of aFPs attributed to \textit{MergirafSemi} may differ when comparing different tools because they successfully resolve different subsets of scenarios.}

\ph{To directly address whether this trade-off holds when restricted to the subset comparable with \textsc{S3M}, we recompute the aFPs and aFNs for \textit{MergirafSemi} and \textit{diff3} on the same 852 scenarios in which \textsc{S3M} completes successfully. Under this restricted setting, \textit{MergirafSemi} yields 1 aFP and 55 aFNs, while \textit{diff3} yields 13 aFPs and 1 aFN. This confirms that the same pattern observed across the full dataset persists on this subset: \textit{MergirafSemi} continues to substantially reduce aFPs relative to \textit{diff3}, at the cost of a much higher aFN count.}

The trade-off is evident when considering structured merging. Inspired by the high \textit{MergirafSemi}'s aFNs number in Table \ref{tab:rq2_afps_afns}, we compared \textit{Mergiraf}'s accuracy against \textit{diff3}. As expected, \textit{Mergiraf} has more aFNs. For instance, structured merging in Python and Rust yields 377 and 432 aFNs, whereas semistructured merging yields 172 and 244. Additionally, \textit{diff3} produces more aFPs compared to \textit{Mergiraf} than \textit{MergirafSemi}, but this is also expected.

This result highlights an important role for semistructured merging: although it increases the risk of silent integration errors compared to conservative line-based tools, it mitigates the substantially higher number of incorrect automatic merges observed in fully structured tools.

\begin{takeaway}
    \ph{\textbf{Answer to RQ2:} Language-specific semistructured merging (\textsc{S3M}) improves correctness by leveraging domain knowledge but is computationally intensive. Language-agnostic semistructured merging (\textit{MergirafSemi}) reduces unnecessary conflicts while maintaining competitive accuracy and better runtime. Line-based merging (\textit{diff3}) is conservative, often exposing conflicts rather than resolving them. However, it may lead to false negatives and over-report conflicts.}
\end{takeaway}

\subsection{RQ3: What is the impact of the commutative configuration
on the accuracy of \textit{MergirafSemi} and \textit{MergirafSemi+}?}
\label{subsec:result_rq3}

We compare \textit{MergirafSemi} and \textit{MergirafSemi+}, which handle Java declaration order differently. \textit{MergirafSemi+} treats method and variable declarations as commutative, allowing conflict-free merges.

Both configurations show similar conflict resolution rates, with \textit{MergirafSemi+} slightly improving over \textit{MergirafSemi} (81.54\% vs. 81.27\%). This indicates that loosening ordering constraints can reduce some unnecessary conflicts. Overall, this specific type of ordering-related conflict seems infrequent, and relaxing constraints has a limited but consistent effect.

A more detailed analysis of merge correctness, measured by the number of aFPs and aFNs, reveals a subtle trade-off introduced by relaxing ordering constraints.

\begin{table}[h!]
\caption{Merge Performance: \textit{MergirafSemi} vs \textit{MergirafSemi+}}
\label{tab:rq3}
\begin{tabular}{lccc}
\toprule
\textbf{Tool}                  & \textbf{Res. Rate} & \textbf{aFPs} & \textbf{aFNs} \\
\midrule
\textit{\textbf{MergirafSemi}}  & 81.27\% & 0 & 0 \\
\textit{\textbf{MergirafSemi+}} & 81.54\% & 0 & 4 \\
\bottomrule
\end{tabular}
\end{table}

In Table \ref{tab:rq3}, both \textit{MergirafSemi} and \textit{MergirafSemi+} produce no additional false positives, indicating that treating declarations as commutative does not introduce false positives. However, \textit{MergirafSemi+} introduces a small number of false negatives (four cases), whereas the default configuration does not introduce any. 

Overall, the results suggest that language-agnostic semistructured merging provides a practical balance between effectiveness and efficiency. While it does not eliminate the trade-offs inherent to language-specific semistructured tools in terms of accuracy, it achieves competitive results while significantly improving upon line-based tools in terms of conflict reduction, without incurring prohibitive runtime costs.

\begin{takeaway}
    \ph{\textbf{Answer to RQ3:} \textit{MergirafSemi+} resolves scenarios that \textit{MergirafSemi} flags as conflicting, showing a trade-off between reducing conflicts and maintaining conservative detection. These cases are few but illustrate more liberal merge decisions. Both configurations} \ph{have almost identical runtime, as the modification only affects local merge decisions. Given minor differences, the choice between \textit{MergirafSemi} and \textit{MergirafSemi+} is a matter of user preference.}
\end{takeaway}
\section{Threats to Validity}
\label{sec:threats_to_validity}

We discuss threats to validity in accordance with standard guidelines for empirical software engineering studies \cite{WOHLIN}.

\textit{Construct Validity.} Our analysis of added false positives (AFPs) and false negatives (AFNs) follows prior work \cite{structured-tools-lastmerge}, combining comparison with repository merges and build/test validation. While useful, these signals are imperfect: repository merges only approximate expected results, and tests are only as strong as the project's test suite. Our tool also depends on Tree-sitter parsers, whose limitations may affect structural representations and merge behavior. To reduce infrastructure-related noise on false negatives, we apply a multi-stage validation process, including CI log inspection and validation of parent commits to distinguish merge-induced failures from external issues. Despite these precautions, the process remains heuristic-based, though it reflects the current gold standard in merge tool evaluation \cite{tools-evaluation-mike-ernst}.

\textit{Internal Validity.} Threats to internal validity concern factors that may affect the correctness of the experimental procedure. Although the evaluation pipeline is fully automated, it relies on custom extensions implemented in a framework \cite{miningframework}. Bugs or misconfigurations in this infrastructure may influence the results. Additionally, differences in tool configuration and execution environments may introduce unintended bias. We mitigate this risk by standardizing execution through Docker containers and applying consistent normalization procedures across all tools.

\textit{External Validity.} Our dataset consists of merge scenarios extracted from publicly available repositories across multiple programming languages. Although this increases diversity compared to single-language studies, it still represents only a subset of possible development contexts. Moreover, the selection criteria focus on non-fast-forward merges involving concurrent modifications, which may not capture all types of merge scenarios encountered in practice. As a result, our findings may not generalize to all repositories, programming languages, or development workflows. The dataset, however, is significant and comparable to the strongest datasets in related work.

\textit{Conclusion Validity.} Our evaluation focuses on scenarios where tools disagree on the presence of conflicts. While this allows us to focus on cases where merge strategies differ, it may not fully capture the tools' overall behavior. Additionally, build and test execution is only performed for these scenarios. As a result, incorrect merges may go undetected when all tools agree. Finally, although execution time is measured using multiple runs to reduce noise, variability in system performance and workload may still influence runtime results.
\section{Related Work}
\label{sec:related_works}

Software merging has been extensively studied, with various tools proposed to improve concurrent change integration. We review existing work, starting with semistructured merge tools, which are mostly related to our proposal, then structured and textual tools.

\subsubsection*{Semistructured Merge}
\label{subsec:semistructured_merge}
\hfill

Semistructured merge tools aim to balance structure-aware merging with the flexibility of textual tools by selectively incorporating syntactic information.

Apel et al. \cite{semistructured-tools-sven} proposed \textit{FSTMerge}, a generic semistructured merge tool based on Feature Structure Trees (FSTs), instantiated for languages such as C\#, Java, and Python. Its core idea of combining structural and textual representations inspires our work. However, \textit{FSTMerge} requires manually annotated grammars, which demand language-specific expertise and limit scalability. In contrast, \textit{MergirafSemi} leverages existing parser infrastructure and lightweight configuration to reduce the effort required to support new languages. Additionally, it compares tools through conflict-resolution metrics. We adopt a more comprehensive evaluation methodology that considers not only conflict resolution but also merge accuracy through compilation and test validation across a broader set of programming languages.

Cavalcanti et al. \cite{semistructured-tools-s3m} introduced \textsc{S3M}, a Java-specific extension of FSTMerge that refines merge results through heuristic handlers, improving accuracy in cases such as renaming. However, it remains tied to a single language. In this work, we compare the two tools to assess whether a language-agnostic semistructured tool can match the performance of language-specific solutions.

Cavalcanti et al. \cite{semistructured-tools-sesame} proposed \textsc{SESAME}, which combines semistructured and textual strategies using language-specific syntactic separators. While it reduces false positives compared to line-based merging, it may increase missed conflicts.

Cavalcanti et al. \cite{structured_impact} directly compared semistructured and structured merge in language-specific settins. Results show that the two strategies differ mainly in conflicting scenarios and exhibit a clear trade-off: semistructured merge reports more false positives, whereas structured merge introduces more false negatives due to more aggressive merge decisions. Our findings align with this observation, observing the same trade-off despite focusing on language-agnostic tools.

Seibt et al. \cite{structure-is-good} compared unstructured to semistructured and structured tools, as well as combinations of them. Results found that increasing structural granularity reduces reported conflicts but raises computational cost and false negatives. They suggest hybrid strategies combine the strengths of both structured and simpler techniques. Our results complement this by showing similar trade-offs.

Tavares et al. \cite{alberto-ase} found semistructured merge offers limited benefits for JavaScript, requiring complex adaptations and yielding only modest conflict reduction. This shows semistructured merge effectiveness depends on language, underscoring the need for flexible designs.

\subsubsection*{Structured Merge}
\label{subsec:structured_merge}
\hfill\

Structured merge tools operate on explicit program representations, typically ASTs, to integrate concurrent changes, more accurately distinguishing independent from interfering modifications than unstructured tools.

Apel et al. \cite{structured-tools-sven} introduced \textit{jDime}, an AST-based merge tool for Java that combines structured and line-based methods through an auto-tuning mechanism. Subsequent work improved its matching through look-ahead strategies and similarity-based heuristics, enhancing alignment and handling refactorings such as renaming and code movement \cite{structured-tools-zhu-jdime}. Despite these advances, \textit{jDime} relies on language-specific parsing, limiting its applicability.

Larsen et al. \cite{structured-tools-spork} proposed \textit{Spork}, which also relies on tree matching (via GumTree \cite{gumtree}) and emphasizes output quality through high-fidelity pretty-printing. While it preserves code structure more effectively than \textit{jDime}, it shares the same limitation of relying on language-specific representations. The core matching and merging algorithms adopted by \textit{Mergiraf} are inspired by \textit{Spork}. As a result, \textit{MergirafSemi} also inherits from this work.

Duarte et al. \cite{structured-tools-lastmerge} introduced \textsc{LastMerge}, a language-agnostic structured merge tool built on Tree-sitter\ph{, with algorithms inspired by jDime}. Their results show that generic structured merging can achieve accuracy and performance comparable to language-specific tools, supporting the feasibility of language-agnostic designs.

\subsubsection*{Textual Merge}
\label{subsec:textual_merge}
\hfill\

Textual merge tools operate directly on raw source code, typically relying on line-based algorithms such as \textit{diff3} to integrate concurrent changes. Because they ignore program syntax, they may produce imprecise merge decisions, motivating tools that refine textual matching.

Clementino et al. \cite{semistructured-tools-csdiff} introduced \textit{CSDiff}, a lightweight tool that enhances line-based merging by incorporating language-aware separators without relying on explicit structural representations. It preprocesses code by isolating syntactic separators (e.g., braces and parentheses) into separate lines, improving alignment across revisions. As an extension, \textit{SepMerge} \cite{semistructured-tools-sepmerge} applies this transformation selectively, refining only conflicting regions identified by \textit{diff3}. However, introducing extra lines can affect LCS computations, potentially leading to unstable results. Additionally, both tools depend on manually defined separator sets, which may not generalize well across languages.

MergeGen \cite{mergegen} uses generative models for conflict resolution, relying on training data rather than explicit program structure. Our tool instead uses deterministic, semistructured merging, but it could be useful for future investigation.

In our design, \textit{MergirafSemi} incorporates line-based merging as a modular component within unstructured regions, allowing it to accommodate and emulate such tools, such as \textit{CSDiff}.
\section{Conclusions}
\label{sec:conclusion}

In this paper, we propose a language-agnostic semistructured merge tool, \textit{MergirafSemi}, balancing structural awareness with language-independent flexibility. Using lightweight syntactic information via CSTs and semistructural reasoning, it avoids the complexity of structured merging and improves on traditional line-based tools.

Our evaluation provides three main findings. First, increasing structural granularity improves conflict resolution and reduces aFPs, but leads to more aggressive merge decisions and a higher risk of aFNs. In contrast, \textit{MergirafSemi} achieves a more balanced trade-off, maintaining competitive accuracy while providing better runtime performance in most common scenarios. Second, when compared to \textit{diff3} and \textsc{S3M}, \textit{MergirafSemi} reduces aFPs relative to \textit{diff3} and achieves comparable accuracy to \textsc{S3M} with no prohibitive runtime overhead, despite being language-agnostic. Third, relaxing context constraints has only a marginal effect on conflict resolution while introducing few aFNs.

Overall, our results show that semistructured merging can be effectively realized in a language-agnostic setting. This positions language-agnostic semistructured merging as a viable alternative to language-agnostic tools, achieving competitive accuracy with language-specific semistructured tools.
\section*{Artifact Availability}
\label{sec:artifact_availability}

To support reproducibility, we provide a replication package containing all artifacts required to reproduce our results \cite{replication_package}. The package includes the dataset of merge scenarios, the exact versions of all evaluated tools (\textit{Mergiraf}, \textit{MergirafSemi}, \textit{MergirafSemi+}, and \textsc{S3M}), the experiment infrastructure, and the scripts used to execute the experiments.
\section*{Acknowledgements}

We thank the participants of our survey and interviews. For partially supporting this work, we would like to thank INES (National Software Engineering Institute) and the Brazilian research funding agencies CNPq, FACEPE, and CAPES.

\bibliographystyle{ACM-Reference-Format}
\bibliography{sample-base}


\end{document}